\documentclass[useAMS,usenatbib]{mn2e}

\def\Msun {\,\mathrm{M}_\odot}

\usepackage{amsmath}
\usepackage{amssymb}
\usepackage{graphicx}
\usepackage{color}
\usepackage{url}
\title[EMP Stars in the TOPoS Survey]{Fingerprint of the first stars: multi-enriched extremely metal-poor stars in the TOPoS survey}

\author[T. Hartwig et al.]
{\parbox{\textwidth}{
Tilman Hartwig$^{1,2}$\thanks{E-mail: tilman.hartwig@ipmu.jp}, Miho N. Ishigaki$^1$, Ralf S. Klessen$^{3,4}$, Naoki Yoshida$^{1,2}$}
\vspace{2mm}\\
$^1$Kavli IPMU (WPI), The University of Tokyo, Kashiwa, Chiba 277-8583, Japan\\
$^2$Department of Physics, School of Science, University of Tokyo, Bunkyo, Tokyo 113-0033, Japan\\
$^3$Universit\"at Heidelberg, Zentrum f\"ur Astronomie, ITA, Albert-Ueberle-Stra{\ss}e 2, 69120 Heidelberg, Germany\\
$^{4}$Universit\"at Heidelberg, Interdisziplin\"ares Zentrum f\"ur Wissenschaftliches Rechnen, INF 205, 69120 Heidelberg, Germany
}

\begin{document}


\pagerange{\pageref{firstpage}--\pageref{lastpage}} \pubyear{2018}

\maketitle

\label{firstpage}

\begin{abstract}
Extremely metal poor (EMP) stars in the Milky Way inherited the chemical composition of the gas out of which they formed. They therefore carry the chemical fingerprint of the first stars in their spectral lines.
It is commonly assumed that EMP stars form from gas that was enriched by only one progenitor supernova (`mono-enriched'). However, recent numerical simulations show that the first stars form in small clusters. Consequently, we expect several supernovae to contribute to the abundances of an EMP star (`multi-enriched').
We analyse seven recently observed EMP stars from the TOPoS survey by applying the divergence of the chemical displacement and find that J1035$+$0641 is mono-enriched ($p_\mathrm{mono}=53\%$) and J1507$+$0051 is multi-enriched ($p_\mathrm{mono}=4\%$). For the remaining five stars we can not make a distinct prediction ($p_\mathrm{mono} \lesssim 50\%$) due to theoretical and observational uncertainties. Further observations in the near-UV will help to improve our diagnostic and therefore contribute to constrain the nature of the first stars.
\end{abstract}


\section{Introduction}
The gas in the early Universe consists only of hydrogen, helium and trace amounts of lithium until the first stars dramatically changed this picture a few hundred million years after the Big Bang. The supernova (SN) explosions of the first, so-called Population~III (Pop~III), stars created the first heavy elements, which enriched the interstellar and intergalactic medium. Primordial gas has lower radiative cooling rates compared to metal-enriched gas and we consequently expect the characteristic clump mass of thermodynamically induced fragmentation in metal-free gas to be higher \citep{omukai05}. Therefore, also the first stars have been suggested to have higher masses, on average, compared to present-day stars \citep{bromm99,abel00}. Despite intensive research, their initial mass function (IMF) is still a matter of ongoing debate. However, their IMF and corresponding feedback efficiency is crucial to understand galaxy formation, reionization, and the formation of the first supermassive black holes.

Extremely metal-poor (EMP) stars in the Milky Way have formed in the early Universe and still carry the characteristic chemical fingerprint of their birth environment. Spectroscopic observations of such EMP stars allow to derive their atmospheric abundances and compare these to the theoretical yields of SN explosion models from metal-free stars. With this approach, several groups have successfully derived mass estimates of individual Pop~III stars \citep{nomoto06,ishigaki14,tominaga14,keller14,ji15,placco15,placco16,fraser17,chen17,ishigaki18}. However, a main assumption of these models is that exactly one Pop~III SN contributes to the metal enrichment of the gas out of which EMP stars form. We have shown in a previous study that also multi-enrichment by several SNe has to be taken into consideration \citep{hartwig18}, in agreement with numerical simulations that also show that Pop~III stars form in small clusters \citep{turk09,stacy10,greif11b,clark11,hirano14,hartwig15b} and previous observations that also suggest multi-enrichment \citep{limongi03}. In this paper, we analyse observed EMP stars and derive the number of supernovae that have enriched the gas out of which they formed.

The Turn-Off Primordial Stars (TOPoS) survey has been designed to detect EMP turn-off stars in the Galactic halo in order to constrain the properties of the first and second generation of stars \citep{topos13}. For these unevolved stars, their atmospheric abundances are unaltered since their formation. Based on a photometric pre-selection with VLT and high-resolution spectroscopic follow-up with UVES, \citet{topos} have recently presented seven new EMP stars with a metallicity\footnote{Defined as $[\mathrm{A}/\mathrm{B}] = \log_{10}(m_\mathrm{A}/m_\mathrm{B})-\log_{10}(m_{\mathrm{A},\odot}/m_{\mathrm{B},\odot})$, where $m_\mathrm{A}$ and $m_\mathrm{B}$ are the abundances of elements A and B and $m_{\mathrm{A},\odot}$ and $m_{\mathrm{B},\odot}$ are the solar abundances of these elements \citep{asplund09}.} of [Fe/H]$\lesssim -3.5$. The total number of observed EMP stars in this metallicity range is only $\lesssim 100$ \citep{saga,abohalima17}.

In this paper, we connect the chemical abundances of the TOPoS stars to the supernova yields of their potential progenitor stars. For the first time, we derive the probability that the gas out of which these EMP stars formed was enriched by only one previous SN and demonstrate that six of the seven stars show signatures of multi-enrichment.

\section{Methodology}
To derive the probabilities of mono-enrichment, we apply two independent techniques, recently advocated by \citet{hartwig18}. To further determine the most likely progenitor of an EMP star, we apply the fitting routine by \citet{ishigaki18}. The theoretical yields of the SN explosions are based on \citet{nomoto13} and for the faint SNe on \citet{ishigaki14}. We do not include lithium in our diagnostics due to still unconstrained lithium depletion mechanism, especially at low metallicities \citep{matsuno17}.

\subsection{Semi-Analytical Model}
We apply a semi-analytical model that is based on \citet{hartwig15} with recent improvements by \citet{magg18}. The model uses MW-like dark matter merger trees from the Caterpillar simulations \citep{griffen16} and determines the formation sites of the first and second generation of stars self-consistently. Chemical and radiative feedback are taken into account and we distinguish between internal and external metal enrichment \citep[for more details, see][]{hartwig18}. After an event of Pop~III star formation in a minihalo we wait for a specific time (of the order $10-100$\,Myr) for the gas to cool and recollapse. Then, we sum all elements that have contributed to the metal enrichment of this halo and trigger the formation of second generation stars.

We do not account for inhomogeneous mixing of the ISM or surface pollution of the stars after their formation, which could lead to systematic differences between the SN yields and the observed atmospheric abundances \citep{yoshii81,hattori14,johnson15,shen16,tanaka17,tanikawa18}. We instead assume that the chemical composition of the star-forming cloud is the same as the composition of the resulting EMP stars and the same as their chemical abundance observed at $z=0$.

We then determine which stars formed out of gas that was enriched by only one SN (`mono-enriched') and which stars formed out of gas that was enriched by more than one SN (`multi-enriched'). The ratio of these two classes of stars as a function of their chemical abundances yields the probability of mono enrichment,
\begin{equation}
\label{eq:pmono}
p_\mathrm{mono} = \frac{N_\mathrm{mono}}{N_\mathrm{mono}+N_\mathrm{multi}},
\end{equation}
where $N_\mathrm{mono}$ and $N_\mathrm{multi}$ are the number of mono- and multi-enriched stars for a certain chemical composition. To account for theoretical and observational uncertainties we convolve this probability distribution with a Gaussian kernel of width $0.5$\,dex \citep{nomoto13,ishigaki18}. Based on our assumptions on star formation in the early Universe, the probability for a second-generation star to be mono-enriched is on average only $\sim 1\%$. However, for EMP stars with metallicities [Fe/H] $\lesssim -3.5$ this probability is on average $p_\mathrm{mono}\geq 20\%$ \citep{hartwig18}.

\subsection{Divergence of the Chemical Displacement}
We briefly describe our new diagnostic based on the divergence of the chemical displacement (DCD) that was first presented in \citet{hartwig18}. The DCD is an analytical estimate based on the theoretical elemental yields of SNe and predicts parameter ranges of elemental abundances that are favourable for mono-enriched stars or ranges that are disfavoured.
 
When two different SNe contribute to the metal enrichment of the interstellar medium (ISM), the resulting elemental abundances are different from the original abundances of each individual SN. This displacement of chemical abundances can be illustrated by two vectors from the yields of the two original SNe to the yields of the resulting composition. The vectors are $N$-dimensional objects for $N$ different elemental abundances with the components of each vector defined as
\begin{equation}
v_i = [\mathrm{X}_i/\mathrm{Fe}]_\mathrm{mix} - [\mathrm{X}_i/\mathrm{Fe}]_\mathrm{SN},
\end{equation}
where $\mathrm{X}_i$ is the abundance of the $i$-th element ($1 \leq i \leq N$) and `SN' and `mix' refer to the abundance ratios of the individual SN and of the mix with a second SN, respectively.

For 25 different Pop~III SNe in the mass range $10-40\Msun$ and $140-150\Msun$ we determine the corresponding $25^2$ vectors of possible SN pair mixing combinations. The upper limit of the mass range has been chosen to reproduce the metallicity distribution function and the fraction of carbon-enhanced metal-poor stars in the MW \citep{hartwig18}. The SN progenitor masses are chosen to sample a logarithmically flat IMF. The total number of 25 results from the constraint that we want $40\%$ of core-collapse supernovae to be represented by at least 8 different faint SNe. We have verified that neither the total number of SNe nor the shape of the IMF qualitatively affect the DCD-based prediction, as long as the desired mass range is covered.

Then, we calculate the divergence of this vector field by applying Gauss integral theorem. The divergence generally quantifies the sinks and sources of a vector field or in terms of our chemical displacement it is positive in regions dominated by the yields of single SNe and negative in regions dominated by the mixture of several SNe. Consequently, a positive DCD is an indicator of mono-enrichment and a negative DCD indicates abundances typical for multi-enrichment. A region of negative divergence can also contain mono-enriched stars, which are difficult to distinguish from multi-enriched ones. For example, the chemical signature of a pair-instability supernova (PISN) with high absolute metal yields is very close to the signature of the same PISN combined with the yields of a faint SN with significantly smaller metal yields. The chemical composition of the resulting enriched ISM is dominated by the PISNe but the stars that form out of this gas have to be considered as multi-enriched. As for the semi-analytical model, we smooth the distribution of the DCD on a scale of $0.5$\,dex to account for theoretical and observational uncertainties.

\begin{figure}
\centering
\includegraphics[width=0.47\textwidth]{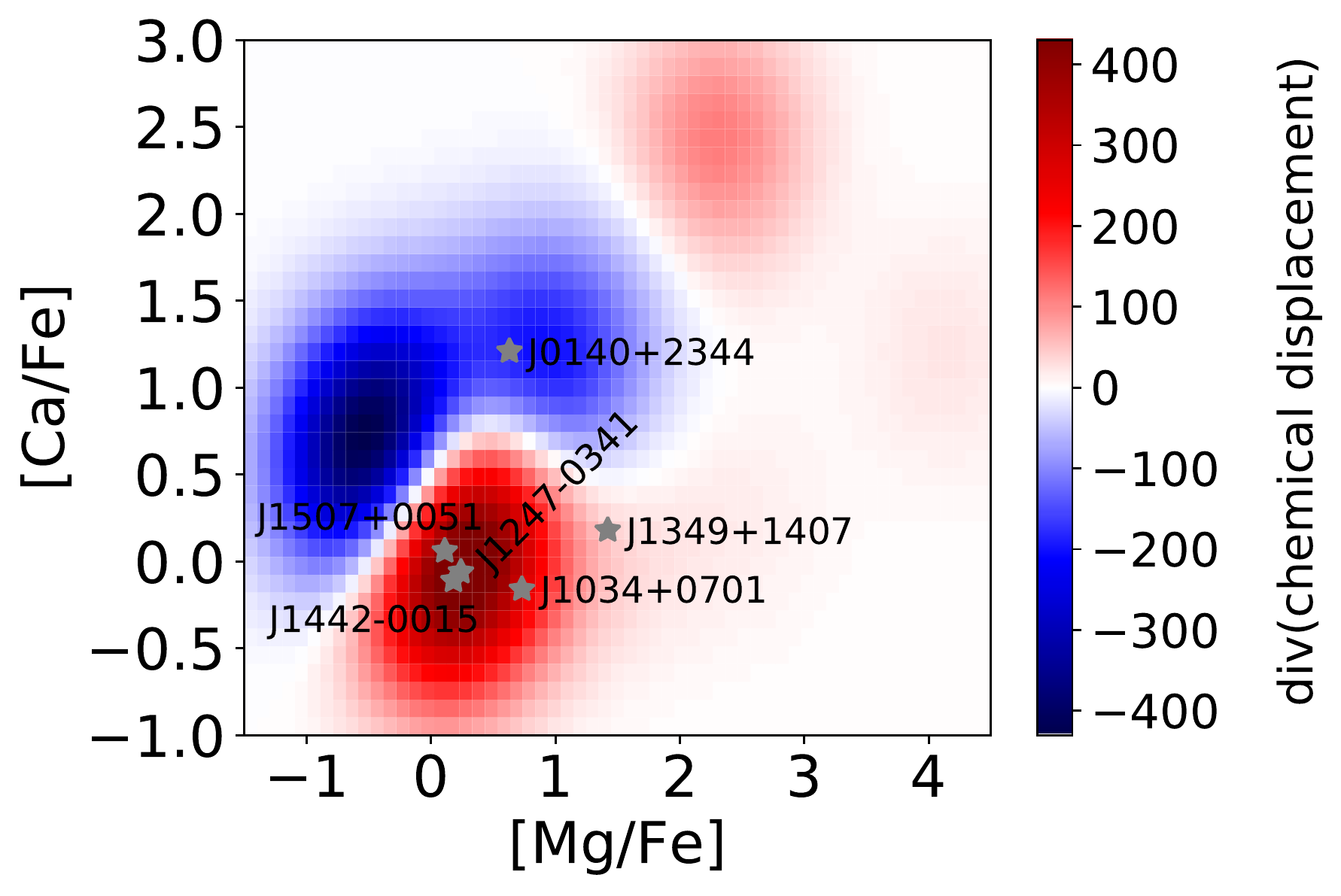}
\caption{DCD as a function of the abundances of calcium and magnesium relative to iron overplotted with the observationally derived values. We see a clear separation into regions with a positive and negative DCD and all but one star have a positive DCD in this representation. J1035+0641 has no magnesium measurement and can therefore not be compared. For the final diagnostic, we include all available elements, which will change the DCD values compared to this example based on only two elements.}
\label{fig:DCD}
\end{figure}

For each observed EMP star the numbers of available elements differ, depending on the quality and wavelength coverage of the spectrum. The divergence has the convenient property that it naturally increases with the number of dimensions, i.e. available elements. An EMP star with more abundances known is therefore expected to show a more pronounced tendency towards positive or negative values of the DCD. We exemplarily show the DCD in two dimensions for calcium and magnesium in Fig. \ref{fig:DCD}.
We chose calcium and magnesium because they are available for all but one star in our sample. Based on this analysis on two elements only, J0140+2344 seems to be multi-enriched and J1349+1407 has a DCD close to zero with little constraining power because its abundances are too far from the input yields from theoretical SNe. The remaining four EMP stars have a positive DCD and are likely to be mono-enriched if we only consider calcium and magnesium. Note that we use all available elements for the final diagnostic and this representation based on two abundance ratios is for illustration purpose only.

The DCD provides an independent, flexible, and computationally more efficient diagnostic compared to the semi-analytical model. It depends only on the assumed yields of the SNe and provides values close to zero, i.e. with little constraining power, if the observed abundances lie in a sparsely sampled region.

%

\subsection{Abundance Fitting}
The observed abundances are fitted with supernova yields of Pop~III stars by the procedure similar to that adopted in \citet{ishigaki18}. 
The Pop~III supernova yield models is searched among five different masses (13, 15, 25, 40, and 100$\Msun$) for the Pop~III progenitor stars. In this paper, we assume that these stars explode as core-collapse supernovae with explosion energies $E_{51}=E/10^{51}\,\mathrm{erg}=0.5$, 1, 10, 30, and 60 respectively. This is analogous to the progenitor mass-explosion energy relation inferred from local supernova observations \citep{nomoto13}.
Pop~III stars with an initial mass of 100$\Msun$ are thought to undergo pulsational pair instability, while neither of their final fates nor metal yields have been well established. We have included the yield model for a 100$\Msun$ Pop~III star which explodes as a core-collapse supernova with high explosion energy ($E_{51}=60$) in order to take into account a possible metal ejection from such explosions \citep{ohkubo09}.

We adopt the mixing-fallback model \citep{umeda02,tominaga07} to calculate the elemental yields with the amount of mixing and the ejected fraction as free parameters. The values of $\chi ^2$ are then calculated with four free parameters: Pop~III mass, outer boundary of the mixing zone, a fraction of ejected mass in the mixing zone, and the hydrogen-mass that dilute the ejected metals.

We have used the [X/H] values obtained by \citet{topos}, and adopt uncertainties in the range $0.1-0.3$\,dex depending on the elements as in \citet{ishigaki18}. For J1035+0641, for which only C and Ca measurements are available, we assume [Ca/Fe] $=0.4$ and thus [Fe/H] $=-5.23$. The assumed [Fe/H] value is consistent with the upper limit ([Fe/H] $<-5.2$) obtained by \citet{topos}. Titanium and scandium are known to be under-produced in 1D calculations of supernova nucleosynthesis compared to those observed in EMP stars \citep{tominaga07,sneden16}. One reason may be that these elements are synthesised through other channels, such as the neutrino process \citep{kobayashi11b} or jetted SNe \citep{tominaga09}. We therefore treat the theoretical yields of titanium and scandium as lower limits.


\section{Results}
\begin{figure}
\centering
\includegraphics[width=0.47\textwidth]{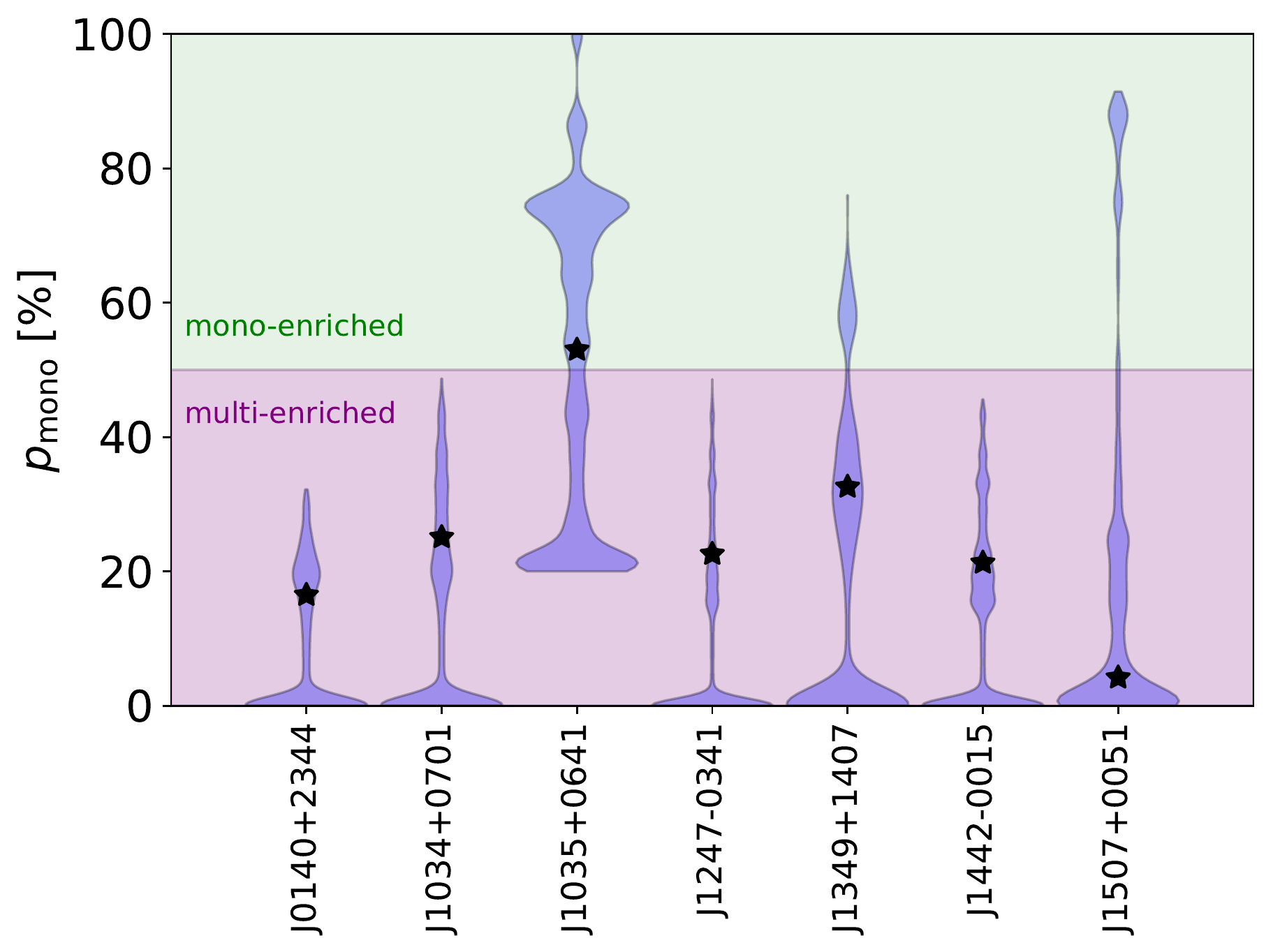}
\caption{Probability distributions for mono-enrichment. The black stars mark the probability for the observationally derived abundances and the blue contours illustrates the probability density for values within $0.5$\,dex of this derived abundance.}
\label{fig:violin}
\end{figure}

We show the probabilities for mono- and multi-enrichment based on the semi-analytical model in Fig.~\ref{fig:violin}.
The vertical axis corresponds to the probability of mono-enrichment (Eq.~\ref{eq:pmono}). The black star illustrates the observationally derived value and the blue contours show the probability distribution within $0.5$\,dex of this value. Six of seven EMP stars are more likely to be multi-enriched according to this analysis. Only the most metal-poor star of this sample, J1035+0641, yields $p_\mathrm{mono}>50\%$. This star has the lowest metallicity in our sample with [Fe/H]$<-5.2$. Interestingly, we do not use this information but only elemental ratios relative to iron. It is rather its high ratio of [C/Fe]$\ge 3.87$ that is a strong indicator for mono-enrichment.

\begin{figure}
\centering
\includegraphics[width=0.47\textwidth]{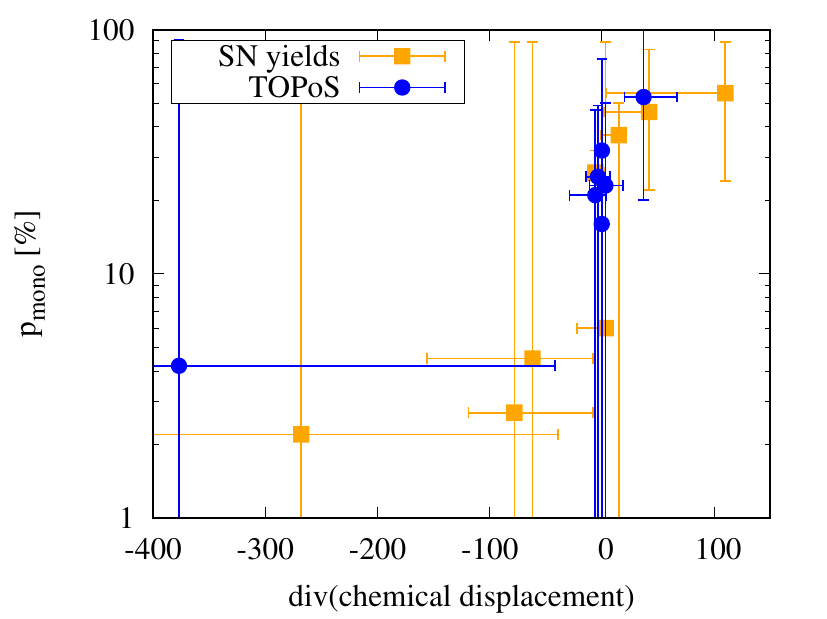}
\caption{Comparison of the DCD with the prediction from the semi-analytical model. The error bars indicate the maximum scatter within $0.5$\,dex of the observed chemical abundance. We slightly offset the values around a divergence of zero for the sake of clarity. The orange squares represent the theoretical yields of different Pop~III SNe (see Tab.~\ref{tab:fiducial}) that were analysed by our diagnostic as if they were observed stars. We confirm the general trend that a negative DCD is an indicator for multi- and a positive value for mono-enrichment. There seems to be an indication of a bimodal distribution, however, the uncertainties in the data together with the small sample size do not allow to draw any firm conclusions.}
\label{fig:compare}
\end{figure}

In Fig.~\ref{fig:compare} we compare the probabilities from the semi-analytical model to the DCD.
We show this comparison for the seven TOPoS EMP stars in blue and for the theoretical SN yields that we have used as input in both models in orange. The probability for mono-enrichment and the DCD are difficult to compare directly due to their very different range of values. However, the data seem to indicate a bimodal distribution of $p_\mathrm{mono}$ with the DCD: for negative values of the DCD we find $1\% \leq p_\mathrm{mono} \leq 5\%$ and for positive values of the DCD we find $30\% \leq p_\mathrm{mono} \leq 50\%$.


Five stars have a DCD close to zero with only little constraining power. This means that their abundances are far from the theoretical model, as it can be seen e.g. from star J1349+1407 in Fig. \ref{fig:DCD}. Only two stars have a DCD that is clearly positive (J1035+0641) or negative (J1507+0051). This demonstrates an advantage of the DCD: it provides values close to zero if the observed abundances are only marginally within the uncertainties of the model. However, this prediction can only be as reliable as the underlying theoretical SN yields and if our model does not cover all variations of different SN this can lead to a DCD close to zero.

None of the orange points corresponds to $p_\mathrm{mono}=100$\%, although this could be expected by construction of the model. This is due to the theoretical and observational uncertainties: even if we observe an EMP star, whose chemical abundances match exactly those of a model Pop~III SN, uncertainties prevent an identification with 100\% reliability. Improvements in the theoretical SN yields and high resolution spectroscopy will help to improve these predictions.

\begin{figure}
\centering
\includegraphics[width=0.47\textwidth]{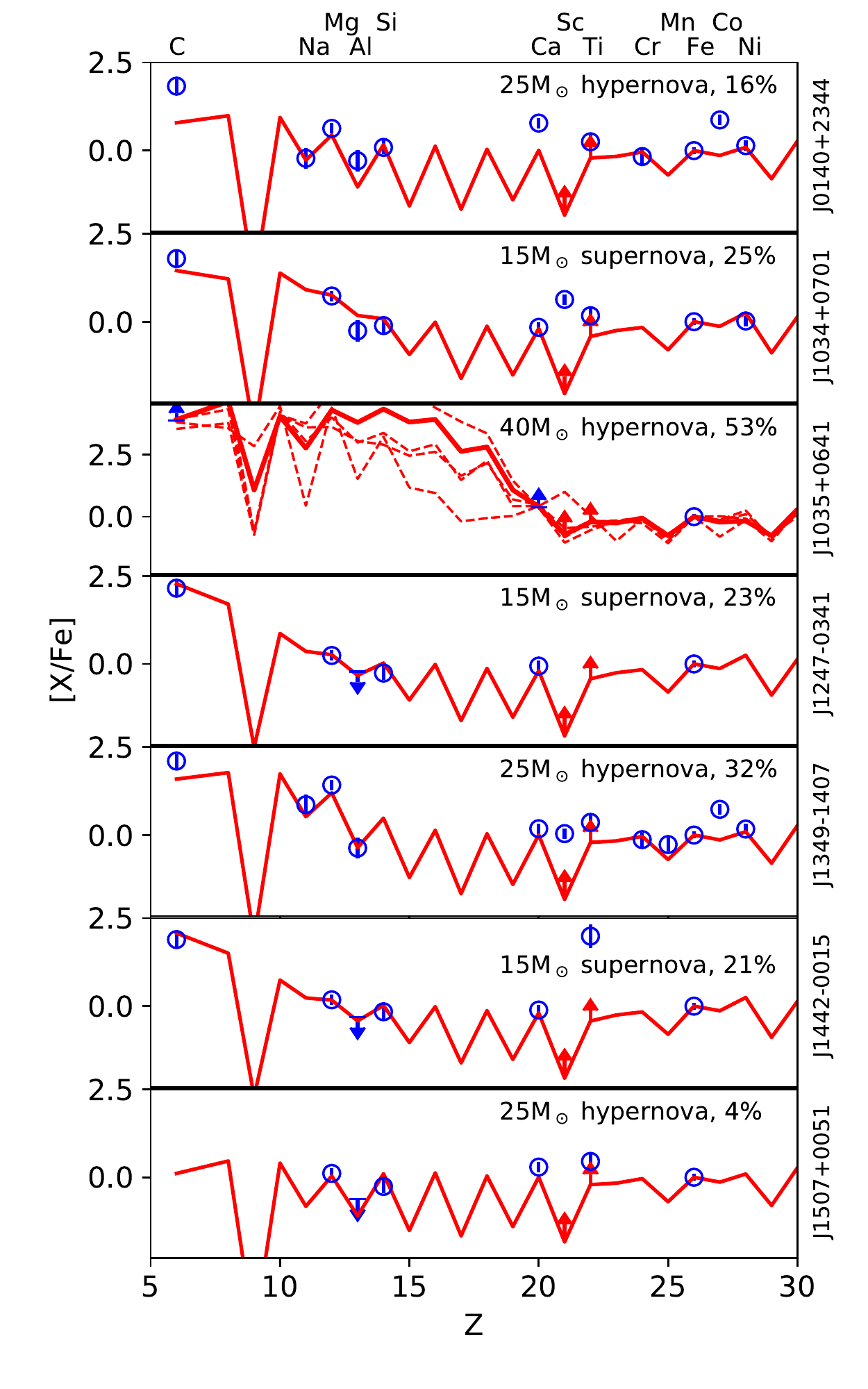}
\caption{Best fits models (red) to the observed abundances (blue). Some models are underconstrained with more free parameters than data points. We illustrate this in the third panel by additionally plotting the next best fitting models. For J1035+0641, we treat the [C/Fe], [Ca/Fe] and [Fe/H] values as measured values rather than upper/lower limits. The most likely progenitor mass and explosion energies, together with the probability of mono-enrichment from the semi-analytical model are provided in the upper right corner of each panel.}
\label{fig:fit}
\end{figure}

Under the assumption that the stars formed out of gas that was enriched by only one progenitor SN, we determine the most likely mass of its Pop~III progenitor star and present the results in Fig~\ref{fig:fit}.
The distribution of progenitor masses and SN explosion energies agrees qualitatively with the results of \citet{ishigaki18}. However, we have a smaller statistical sample and find that six of the seven stars are likely to be multi-enriched and their derived progenitor properties have therefore to be treated with caution.

In Table~\ref{tab:fiducial} we summarize and compare the results for all three diagnostics.
\begin{table*}
 \centering
 \begin{tabular}{lllllllllllll}
 & & available & \multicolumn{3}{c}{prob. mono-enrichment} & \multicolumn{3}{c}{div(chemical displacement)} & progenitor & &E$_\mathrm{SN}$\\ 
 & [Fe/H] & elements & min & best & max & min & best & max & mass & $\chi ^2/$DoF &$10^{51}$\,erg\\ 
 \hline 
 J0140+2344 & $-4.0$ & 12 & $0\%$ & $16\%$ & $32\%$ & $-0.57$ & $-0.18$ & $0.33$ & $25 \Msun $ & $9.6$ & $10$\\ 
 J1034+0701 & $-4.0$ & 9 & $0\%$ & $25\%$ & $49\%$ & $-14$ & $-1.7$ & $7.0$ & $15\Msun$ & $1.2$ & $1$\\ 
 J1035+0641 & $<5.2$ & 3 & $20\%$ & $53\%$ & $100\%$ & $20$ & $37$ & $67$ & $40\Msun$ & DoF$< 0$ & 10\\ 
 J1247$-$0341 & $-4.0$ & 6 & $0\%$ & $23\%$ & $49\%$ & $-11$ & $1.1$ & $19$ & $15\Msun$ & $2.8$ & $1$\\ 
 J1349+1407 & $-3.6$ & 12 & $0\%$ & $32\%$ & $76\%$ & $-0.16$ & $-0.12$ & $1.7$ & $25\Msun$ & $4.8$ & $10$\\ 
 J1442$-$0015 & $-4.4$ & 7 & $0\%$ & $21\%$ & $47\%$ & $-29$ & $-5.9$ & $3.6$ & $15\Msun$ & $1.1$ & $1$\\ 
 J1507+0051 & $-3.4$ & 7 & $0\%$ & $4.2\%$ & $91\%$ & $-736$ & $-377$ & $-42$ & $25\Msun$ & DoF$= 0$ & $10$\\ 
faintSN1 & -- & 12 & $0\%$ & $6.0\%$ & $89\%$ & $-22$ & $3.2$ & $3.2$ & -- & -- & $1$ \\ 
faintSN2 & -- & 12 & $24\%$ & $55\%$ & $89\%$ & $3.3$ & $110$ & $110$ & -- & -- & $1$ \\  
faintSN3 & -- & 12 & $0\%$ & $37\%$ & $50\%$ & $-1.2$ & $15$ & $15$ & -- & -- & $1$ \\  
faintSN4 & -- & 12 & $22\%$ & $46\%$ & $83\%$ & $1.4$ & $42$ & $42$ & -- & -- & $1$ \\  
CCSN15 & -- & 12 & $0.1\%$ & $4.2\%$ & $89\%$ & $-156$ & $-62$ & $-8.2$ & -- & -- & $1$ \\  
CCSN25 & -- & 12 & $0.1\%$ & $2.2\%$ & $89\%$ & $-435$ & $-268$ & $-39$ & -- & -- & $1$ \\  
CCSN40 & -- & 12 & $0.1\%$ & $2.7\%$ & $89\%$ & $-119$ & $-78$ & $-8.3$ & -- & -- & $1$ \\
PISN150 & -- & 12 & $0.1\%$ & $26\%$ & $32\%$ & $-6.9$ & $-6.0$ & $2.8$ & -- & -- & $16$ \\
 \end{tabular} 
  \caption{Summary and comparison of the properties of the seven observed EMP stars from the TOPoS survey and eight different SN models, which were treated as if they were observed abundances. The columns `best' show the values for the exact elemental abundances and the `min' and `max' columns provide the range within $0.5$\,dex of these abundances. The column $\chi ^2 /\mathrm{DoF}$ provides the reduced $\chi ^2$ as a quantification of the fit quality. For two stars the number of degrees of freedom is DoF$\leq 0$ and the corresponding results of the fitting has to be interpreted with caution.}
   \label{tab:fiducial}
\end{table*}
J1035+0641 with the lowest metallicity of this sample has the highest probability for mono enrichment and J1507+0051 with the highest metallicity has the lowest probability. Although such a correlation is expected \citep{hartwig18} the other five stars do not follow such a monotonous trend between the metallicity and the likelihood for mono-enrichment .

\section{Discussion}
The majority of stars in this sample is most likely to be multi-enriched, which allows speculations about the multiplicity of the first stars. A natural explanation for the dominance of multi-enriched EMP stars is that the first stars form in clusters, as advocated by various numerical simulations \citep{turk09,stacy10,greif11b,clark11,hirano14,hartwig15b}. We have chosen to analyse the EMP stars from the TOPoS survey due to their recency and the small sample allows to test, calibrate, and improve our diagnostic. It is a homogeneous sample of lowest metallicity turn-off stars, which are unevolved stars and therefore their atmospheric abundances are less likely affected by internal mixing. Thus, they are most suitable for the purpose of examining chemical enrichments purely by the first stars.
However, the statistics of this sample is very small with only one star that has a clear positive DCD, indicating mono-enrichment. Moreover, other cosmic scenarios for multi-enrichment are possible, such as EMP star formation after the merger of two haloes that both experienced enrichment by a single SN. A future analysis with an improved diagnostic and better statistics is required to constrain the multiplicity of the first stars based on multi-enriched EMP stars.

Our prediction can only be as reliable as the underlying theoretical model for the nucleosynthetic yields of Pop~III SN explosions. There are differences of $\gtrsim 0.5$\,dex for the elemental ratios relative to iron for independent models in the literature \citep{tominaga07,hw10,limongi12,nomoto13}. We include this theoretical together with the observational uncertainty as a constant smoothing kernel. We do not include other enrichment channels in our model, such as mass transfer from AGB stars or neutron star mergers, which is justified because there is no measurement for s- or r-process elements in the surface abundances of these stars, which would require such additional enrichment channels. Since we do not cover all possible enrichment channels, the probability of multi-enrichment, $p_\mathrm{multi}=1-p_\mathrm{mono}$, should therefore rather be interpreted as the probability for not being mono-enriched by Pop~III SNe. Moreover, for the calculation of the DCD we only use 25 different SN, which may not cover all possible variations of SNe regarding explosion energies, asymmetry, and mixing efficiencies.



\subsection{Most informative elements}
\begin{figure}
\centering
\includegraphics[width=0.47\textwidth]{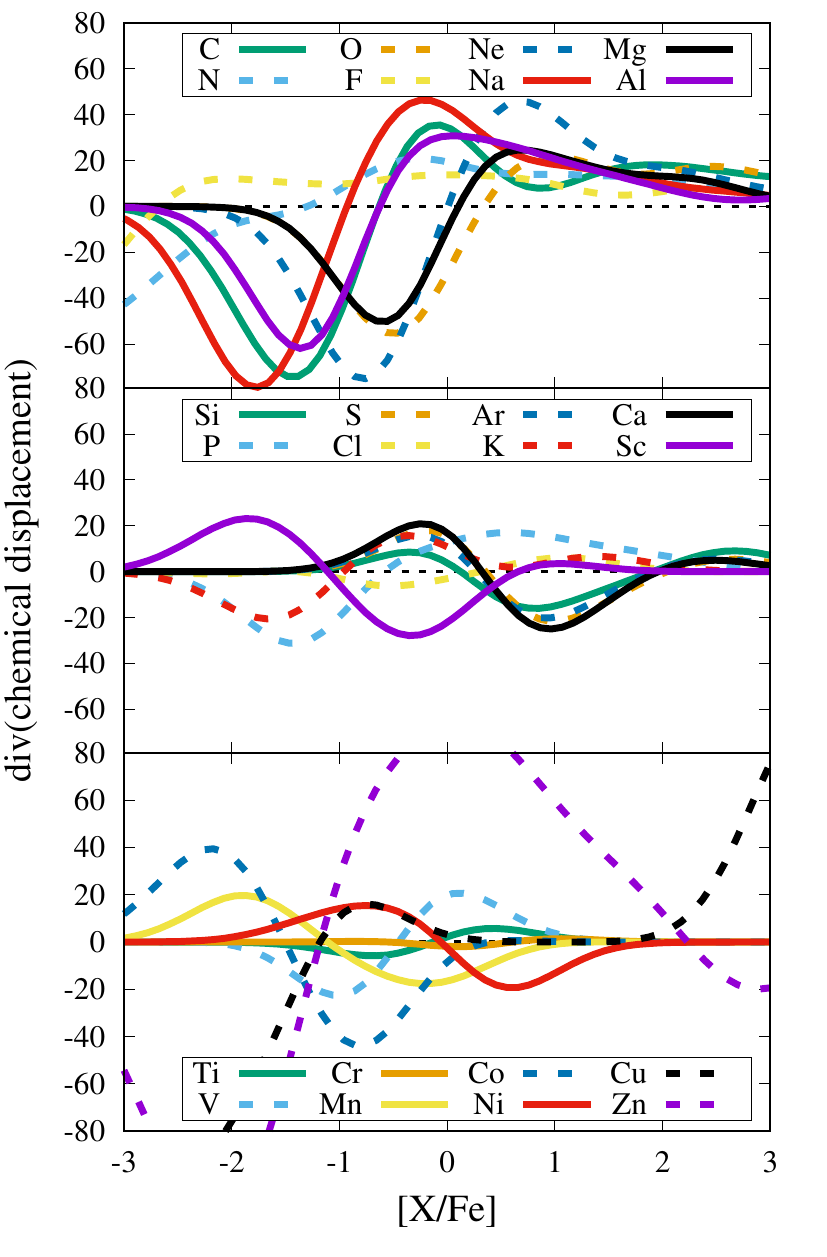}
\caption{DCD for all elements from carbon to zinc relative to iron. Elements that are detected in at least one of the seven EMP stars from the TOPoS survey are shown in solid, other elements with dashed lines. This representation illustrates the most informative elements in order to distinguish mono- from multi-enriched EMP stars. High absolute values and far separated extrema are beneficial for this diagnostic, as for sodium, carbon, or aluminium. Additional observed abundances of e.g. neon, copper, or zinc can further help to improve the diagnostic.}
\label{fig:allDCD}
\end{figure}

In addition to its diagnostic power for observed EMP stars, the DCD can also be used as a tool to predict the constraining power of observed elements. We therefore calculate the DCD in one dimension for various elements and compare their predictive power in Fig.~\ref{fig:allDCD}.
Amongst the elements that are already available for at least some of the seven EMP stars from the TOPoS survey, sodium, carbon, and aluminium provide the highest and lowest values of the DCD and are therefore the most informative elements. We confirm earlier results that carbon, magnesium, and their ratio contain valuable information to discriminate mono- from multi-enriched stars \citep{hartwig18,placco18}. Future observations of these elements will help to obtain more reliable estimates on the enrichment history of these stars.

Silicon, titanium, and chromium, on the other hand, do generally not provide any additional information because the minima and maxima of their 1-dimensional DCD are neither very pronounced nor far separated. This is because the SNe in our model produce all very similar yields of these elements relative to iron and it is therefore difficult to discriminate mono- from multi-enrichment on their basis. This is valid in 1D, corresponding to a situation where only the ratio of this element with respect to iron is available. If additional elemental lines are detected, elements such as silicon and titanium can be used as orthogonal estimate to further improve the diagnostic.

For the elements that are not yet available for these seven EMP stars, we identify neon, copper, and zinc as the most informative and promising elements. Also fluorine and nitrogen can be very helpful because they have a positive DCD for the large range of $-2 \lesssim $[F/Fe]$\lesssim 3$ and $-1 \lesssim $[N/Fe]$\lesssim 3$, respectively.

\subsection{Observational prospects}
The TOPoS survey focuses on turn-off stars, which are brighter compared to the majority of other Galactic halo stars. While this allows to survey also rare populations of stars in sufficient number, the high effective temperatures result in weaker metal absorption lines, compared to cooler stars. Therefore, the detection of elemental abundances with sufficiently high signal-to-noise ratios is more challenging. For the seven EMP stars from the TOPoS survey, an additional measurement of sodium for five of the stars, a measurement of magnesium for J1035$+$0641, and of carbon for J1507$+$0051 would further improve our diagnostics.

Copper, zinc, and nitrogen are promising additional elements to better constrain the nature of EMP stars and consequently of the first stars in the Universe with ground-based observations. The Cu{\sc i} UV resonance lines at $324.7$\,nm and $327.3$\,nm are strong and have been measured in CD$-$38\,245 providing log(Cu/H)$+12 = 0.04\pm0.10$ \citep{andrievsky18}. The Zn abundance has been measured for EMP stars down to metallicities of [Fe/H]$\approx -4$ \citep{cayrel04,bonifacio12}, using the Zn{\sc i} line at 481\,nm. Nitrogen can be traced down to very low abundances using the UV NH band at 336\,nm \citep{collet06}. On the contrary, neon and fluorine, which we also identify as very informative elements, are very difficult to detect, especially at the relevant low abundances for EMP stars \citep{li13}.

The most promising absorption lines that we identify to help further discriminating between mono- and multi-enriched stars are in the near-UV ($\sim 300-500$\,nm). The most informative elements, copper and zinc, can be observed with e.g. the Space Telescope Imaging Spectrograph on board the Hubble Space Telescope \citep{roederer18} with phosphorus and sulphur in the same spectral range \citep{roederer16}. Future spectrographs in the UV to detect these elements are e.g. CUBES \citep{barbuy14}, the Ultraviolet Multi-Object Spectrograph \citep[LUMOS, ][]{france17} on board the future LUVOIR space telescope, or the UV spectrographs on HabEx or CETUS, two further mission in their concept planning phase \citep{habex16,cetus17}.

\section{Summary and Conclusion}
We present formation scenarios for seven recently observed EMP stars from the TOPoS survey \citep{topos}. Due to their low metallicity, the stars are very likely to have formed in the early Universe, after only one previous generation of Pop~III star formation. We interpret the chemical fingerprint in their stellar atmospheres as signature of the first stars and derive the properties of their progenitors.

We provide for the first time probabilities of mono-enrichment. Although most numerical simulations of primordial star formation find that Pop~III stars form in clusters and should give rise to several SNe per minihalo, most abundance fitting models of EMP stars assume only the yields of one progenitor SN. Six of seven EMP stars are likely to be multi-enriched and their derived progenitor properties need to be interpreted by including the corresponding probability for multi-enrichment.

For individual stars we find that very likely J1035+0641 is mono-enriched and J1507+0051 is multi-enriched. This conclusion is independently derived with two different diagnostics: first, based on a semi-analytical approach that models the formation of the first and second generation of stars based on MW-like dark matter merger trees. And second, based on the DCD, which shows clear trends, even within the uncertainty margins. These two stars have the lowest and highest metallicity in the sample. However, the evolution of the probability for mono-enrichment with the metallicity is not monotonous.

We identify the most informative elements to discriminate mono- from multi-enrichment. Sodium, carbon, and aluminium provide the highest information gain among the elements that are already available for some of the seven stars. Additional measurements for copper, zinc, and nitrogen in the near UV will improve the model prediction.

We demonstrate that the likelihood for mono-enrichment can be independently derived with the divergence of the chemical displacement, first proposed in \citet{hartwig18}. This new diagnostics is computationally cheaper, more flexible, and requires less assumptions on the physical model of star formation in the early Universe than the semi-analytical model. A positive DCD corresponds to $p_\mathrm{mono} \gtrsim 20$\%. We can never be sure to 100\% that an EMP star is mono-enriched. Even if we observe abundances that exactly match predicted elemental yields from a Pop~III SNe the observational and theoretical uncertainties prevent such a strong conclusion. Future improvements on the theoretical models of SN yields and on observationally derived abundances can minimise this effect and allow for tighter constraints.


We derive the predictive power of two independent diagnostics to discriminate between mono- and multi-enriched EMP stars, based on their elemental abundances. Additional observations, especially in the near-UV, can help to further improve the constraints. In a future study, we plan to apply this diagnostic with improved theoretical SN yields to a larger sample of EMP stars to derive statistically sound predictions on the nature of the first stars.

\subsection*{Acknowledgements}
We appreciate valuable comments on the manuscript from Piercarlo Bonifacio, Elisabetta Caffau, and Ian Roederer, especially on the observational perspectives. We thank the reviewer for constructive suggestions and careful reading of the manuscript.
TH is a JSPS International Research Fellow. RSK was supported by the European Research Council under the European Community's Seventh Framework Programme (FP7/2007 - 2013) via the ERC Advanced Grant `STARLIGHT: Formation of the First Stars’ under the project number 339177 and by the Deutsche Forschungsgemeinschaft via SFB 881 `The Milky Way System' (subprojects B1, B2, and B8) and SPP 1573 `Physics of the Interstellar Medium' (grant numbers KL 1358/18.1, KL 1358/19.2 and GL 668/2-1). This work was supported by World Premier International Research Center Initiative (WPI Initiative), MEXT, Japan. We are very grateful to the Caterpillar collaboration for providing their dark matter merger trees.

\bibliographystyle{mn2e}
\bibliography{topos}

\begin{thebibliography}{}

\bibitem[\protect\citeauthoryear{{Abel}, {Bryan} \& {Norman}}{{Abel}
  et~al.}{2000}]{abel00}
{Abel} T.,  {Bryan} G.~L.,    {Norman} M.~L.,  2000, \apj, 540, 39

\bibitem[\protect\citeauthoryear{{Abohalima} \& {Frebel}}{{Abohalima} \&
  {Frebel}}{2017}]{abohalima17}
{Abohalima} A.,  {Frebel} A.,  2017, ArXiv e-prints 1711.04410

\bibitem[\protect\citeauthoryear{{Andrievsky}, {Bonifacio}, {Caffau},
  {Korotin}, {Spite}, {Spite}, {Sbordone} \& {Zhukova}}{{Andrievsky}
  et~al.}{2018}]{andrievsky18}
{Andrievsky} S.,  {Bonifacio} P.,  {Caffau} E.,  {Korotin} S.,  {Spite} M.,
  {Spite} F.,  {Sbordone} L.,    {Zhukova} A.~V.,  2018, \mnras, 473, 3377

\bibitem[\protect\citeauthoryear{{Asplund}, {Grevesse}, {Sauval} \&
  {Scott}}{{Asplund} et~al.}{2009}]{asplund09}
{Asplund} M.,  {Grevesse} N.,  {Sauval} A.~J.,    {Scott} P.,  2009, \araa, 47,
  481

\bibitem[\protect\citeauthoryear{{Barbuy} et~al.,}{{Barbuy}
  et~al.}{2014}]{barbuy14}
{Barbuy} B.,  et~al., 2014, \apss, 354, 191

\bibitem[\protect\citeauthoryear{{Bonifacio}, {Caffau}, {Spite}, {Spite},
  {Sbordone}, {Monaco}, {Fran{\c c}ois}, {Plez}, {Molaro}, {Gallagher},
  {Cayrel}, {Christlieb}, {Klessen}, {Koch}, {Ludwig}, {Steffen}, {Zaggia} \&
  {Abate}}{{Bonifacio} et~al.}{2018}]{topos}
{Bonifacio} P.,  {Caffau} E.,  {Spite} M.,  {Spite} F.,  {Sbordone} L.,
  {Monaco} L.,  {Fran{\c c}ois} P.,  {Plez} B.,  {Molaro} P.,  {Gallagher}
  A.~J.,  {Cayrel} R.,  {Christlieb} N.,  {Klessen} R.~S.,  {Koch} A.,
  {Ludwig} H.-G.,  {Steffen} M.,  {Zaggia} S.,    {Abate} C.,  2018, \aap, 612,
  A65

\bibitem[\protect\citeauthoryear{{Bonifacio}, {Caffau}, {Venn} \&
  {Lambert}}{{Bonifacio} et~al.}{2012}]{bonifacio12}
{Bonifacio} P.,  {Caffau} E.,  {Venn} K.~A.,    {Lambert} D.~L.,  2012, \aap,
  544, A102

\bibitem[\protect\citeauthoryear{{Bromm}, {Coppi} \& {Larson}}{{Bromm}
  et~al.}{1999}]{bromm99}
{Bromm} V.,  {Coppi} P.~S.,    {Larson} R.~B.,  1999, \apjl, 527, L5

\bibitem[\protect\citeauthoryear{{Caffau} et~al.,}{{Caffau}
  et~al.}{2013}]{topos13}
{Caffau} E.,  et~al., 2013, \aap, 560, A71

\bibitem[\protect\citeauthoryear{{Cayrel}, {Depagne}, {Spite}, {Hill}, {Spite},
  {Fran{\c c}ois}, {Plez}, {Beers}, {Primas}, {Andersen}, {Barbuy},
  {Bonifacio}, {Molaro} \& {Nordstr{\"o}m}}{{Cayrel} et~al.}{2004}]{cayrel04}
{Cayrel} R.,  {Depagne} E.,  {Spite} M.,  {Hill} V.,  {Spite} F.,  {Fran{\c
  c}ois} P.,  {Plez} B.,  {Beers} T.,  {Primas} F.,  {Andersen} J.,  {Barbuy}
  B.,  {Bonifacio} P.,  {Molaro} P.,    {Nordstr{\"o}m} B.,  2004, \aap, 416,
  1117

\bibitem[\protect\citeauthoryear{{Chen}, {Heger}, {Whalen}, {Moriya}, {Bromm}
  \& {Woosley}}{{Chen} et~al.}{2017}]{chen17}
{Chen} K.-J.,  {Heger} A.,  {Whalen} D.~J.,  {Moriya} T.~J.,  {Bromm} V.,
  {Woosley} S.~E.,  2017, \mnras, 467, 4731

\bibitem[\protect\citeauthoryear{{Clark}, {Glover}, {Smith}, {Greif}, {Klessen}
  \& {Bromm}}{{Clark} et~al.}{2011}]{clark11}
{Clark} P.~C.,  {Glover} S.~C.~O.,  {Smith} R.~J.,  {Greif} T.~H.,  {Klessen}
  R.~S.,    {Bromm} V.,  2011, Science, 331, 1040

\bibitem[\protect\citeauthoryear{{Collet}, {Asplund} \& {Trampedach}}{{Collet}
  et~al.}{2006}]{collet06}
{Collet} R.,  {Asplund} M.,    {Trampedach} R.,  2006, \apjl, 644, L121

\bibitem[\protect\citeauthoryear{{France}, {Fleming}, {West}, {McCandliss},
  {Bolcar}, {Harris}, {Moustakas}, {O'Meara}, {Pascucci}, {Rigby},
  {Schiminovich} \& {Tumlinson}}{{France} et~al.}{2017}]{france17}
{France} K.,  {Fleming} B.,  {West} G.,  {McCandliss} S.~R.,  {Bolcar} M.~R.,
  {Harris} W.,  {Moustakas} L.,  {O'Meara} J.~M.,  {Pascucci} I.,  {Rigby} J.,
  {Schiminovich} D.,    {Tumlinson} J.,  2017, in Society of Photo-Optical
  Instrumentation Engineers (SPIE) Conference Series Vol. 10397 of Society of
  Photo-Optical Instrumentation Engineers (SPIE) Conference Series, {The LUVOIR
  Ultraviolet Multi-Object Spectrograph (LUMOS): instrument definition and
  design}.
p. 1039713

\bibitem[\protect\citeauthoryear{{Fraser}, {Casey}, {Gilmore}, {Heger} \&
  {Chan}}{{Fraser} et~al.}{2017}]{fraser17}
{Fraser} M.,  {Casey} A.~R.,  {Gilmore} G.,  {Heger} A.,    {Chan} C.,  2017,
  \mnras, 468, 418

\bibitem[\protect\citeauthoryear{{Greif}, {Springel}, {White}, {Glover},
  {Clark}, {Smith}, {Klessen} \& {Bromm}}{{Greif} et~al.}{2011}]{greif11b}
{Greif} T.~H.,  {Springel} V.,  {White} S.~D.~M.,  {Glover} S.~C.~O.,  {Clark}
  P.~C.,  {Smith} R.~J.,  {Klessen} R.~S.,    {Bromm} V.,  2011, \apj, 737, 75

\bibitem[\protect\citeauthoryear{{Griffen}, {Ji}, {Dooley}, {G{\'o}mez},
  {Vogelsberger}, {O'Shea} \& {Frebel}}{{Griffen} et~al.}{2016}]{griffen16}
{Griffen} B.~F.,  {Ji} A.~P.,  {Dooley} G.~A.,  {G{\'o}mez} F.~A.,
  {Vogelsberger} M.,  {O'Shea} B.~W.,    {Frebel} A.,  2016, \apj, 818, 10

\bibitem[\protect\citeauthoryear{{Hartwig}, {Bromm}, {Klessen} \&
  {Glover}}{{Hartwig} et~al.}{2015}]{hartwig15}
{Hartwig} T.,  {Bromm} V.,  {Klessen} R.~S.,    {Glover} S.~C.~O.,  2015,
  \mnras, 447, 3892

\bibitem[\protect\citeauthoryear{{Hartwig}, {Clark}, {Glover}, {Klessen} \&
  {Sasaki}}{{Hartwig} et~al.}{2015}]{hartwig15b}
{Hartwig} T.,  {Clark} P.~C.,  {Glover} S.~C.~O.,  {Klessen} R.~S.,    {Sasaki}
  M.,  2015, \apj, 799, 114

\bibitem[\protect\citeauthoryear{{Hartwig}, {Yoshida}, {Magg}, {Frebel},
  {Glover}, {G{\'o}mez}, {Griffen}, {Ishigaki}, {Ji}, {Klessen}, {O'Shea} \&
  {Tominaga}}{{Hartwig} et~al.}{2018}]{hartwig18}
{Hartwig} T.,  {Yoshida} N.,  {Magg} M.,  {Frebel} A.,  {Glover} S.~C.~O.,
  {G{\'o}mez} F.~A.,  {Griffen} B.,  {Ishigaki} M.~N.,  {Ji} A.~P.,  {Klessen}
  R.~S.,  {O'Shea} B.~W.,    {Tominaga} N.,  2018, \mnras, 478, 1795

\bibitem[\protect\citeauthoryear{{Hattori}, {Yoshii}, {Beers}, {Carollo} \&
  {Lee}}{{Hattori} et~al.}{2014}]{hattori14}
{Hattori} K.,  {Yoshii} Y.,  {Beers} T.~C.,  {Carollo} D.,    {Lee} Y.~S.,
  2014, \apj, 784, 153

\bibitem[\protect\citeauthoryear{{Heap} \& {CETUS Team}}{{Heap} \& {CETUS
  Team}}{2017}]{cetus17}
{Heap} S.~R.,  {CETUS Team} 2017, in American Astronomical Society Meeting
  Abstracts \#229 Vol.~229 of American Astronomical Society Meeting Abstracts,
  {Cosmic Evolution Through UV Spectroscopy (CETUS): A NASA Probe-Class Mission
  Concept}.
p. 238.27

\bibitem[\protect\citeauthoryear{{Heger} \& {Woosley}}{{Heger} \&
  {Woosley}}{2010}]{hw10}
{Heger} A.,  {Woosley} S.~E.,  2010, \apj, 724, 341

\bibitem[\protect\citeauthoryear{{Hirano}, {Hosokawa}, {Yoshida}, {Umeda},
  {Omukai}, {Chiaki} \& {Yorke}}{{Hirano} et~al.}{2014}]{hirano14}
{Hirano} S.,  {Hosokawa} T.,  {Yoshida} N.,  {Umeda} H.,  {Omukai} K.,
  {Chiaki} G.,    {Yorke} H.~W.,  2014, \apj, 781, 60

\bibitem[\protect\citeauthoryear{{Ishigaki}, {Tominaga}, {Kobayashi} \&
  {Nomoto}}{{Ishigaki} et~al.}{2014}]{ishigaki14}
{Ishigaki} M.~N.,  {Tominaga} N.,  {Kobayashi} C.,    {Nomoto} K.,  2014,
  \apjl, 792, L32

\bibitem[\protect\citeauthoryear{{Ishigaki}, {Tominaga}, {Kobayashi} \&
  {Nomoto}}{{Ishigaki} et~al.}{2018}]{ishigaki18}
{Ishigaki} M.~N.,  {Tominaga} N.,  {Kobayashi} C.,    {Nomoto} K.,  2018, \apj,
  857, 46

\bibitem[\protect\citeauthoryear{{Ji}, {Frebel} \& {Bromm}}{{Ji}
  et~al.}{2015}]{ji15}
{Ji} A.~P.,  {Frebel} A.,    {Bromm} V.,  2015, \mnras, 454, 659

\bibitem[\protect\citeauthoryear{{Johnson}}{{Johnson}}{2015}]{johnson15}
{Johnson} J.~L.,  2015, \mnras, 453, 2771

\bibitem[\protect\citeauthoryear{{Keller} et~al.,}{{Keller}
  et~al.}{2014}]{keller14}
{Keller} S.~C.,  et~al., 2014, \nat, 506, 463

\bibitem[\protect\citeauthoryear{{Kobayashi}, {Izutani}, {Karakas}, {Yoshida},
  {Yong} \& {Umeda}}{{Kobayashi} et~al.}{2011}]{kobayashi11b}
{Kobayashi} C.,  {Izutani} N.,  {Karakas} A.~I.,  {Yoshida} T.,  {Yong} D.,
  {Umeda} H.,  2011, \apjl, 739, L57

\bibitem[\protect\citeauthoryear{{Li}, {Ludwig}, {Caffau}, {Christlieb} \&
  {Zhao}}{{Li} et~al.}{2013}]{li13}
{Li} H.~N.,  {Ludwig} H.-G.,  {Caffau} E.,  {Christlieb} N.,    {Zhao} G.,
  2013, \apj, 765, 51

\bibitem[\protect\citeauthoryear{{Limongi} \& {Chieffi}}{{Limongi} \&
  {Chieffi}}{2012}]{limongi12}
{Limongi} M.,  {Chieffi} A.,  2012, \apjs, 199, 38

\bibitem[\protect\citeauthoryear{{Limongi}, {Chieffi} \& {Bonifacio}}{{Limongi}
  et~al.}{2003}]{limongi03}
{Limongi} M.,  {Chieffi} A.,    {Bonifacio} P.,  2003, \apjl, 594, L123

\bibitem[\protect\citeauthoryear{{Magg}, {Hartwig}, {Agarwal}, {Frebel},
  {Glover}, {Griffen} \& {Klessen}}{{Magg} et~al.}{2018}]{magg18}
{Magg} M.,  {Hartwig} T.,  {Agarwal} B.,  {Frebel} A.,  {Glover} S.~C.~O.,
  {Griffen} B.~F.,    {Klessen} R.~S.,  2018, \mnras, 473, 5308

\bibitem[\protect\citeauthoryear{{Matsuno}, {Aoki}, {Suda} \& {Li}}{{Matsuno}
  et~al.}{2017}]{matsuno17}
{Matsuno} T.,  {Aoki} W.,  {Suda} T.,    {Li} H.,  2017, \pasj, 69, 24

\bibitem[\protect\citeauthoryear{{Mennesson} et~al.,}{{Mennesson}
  et~al.}{2016}]{habex16}
{Mennesson} B.,  et~al., 2016, in Space Telescopes and Instrumentation 2016:
  Optical, Infrared, and Millimeter Wave Vol.~9904 of \procspie, {The Habitable
  Exoplanet (HabEx) Imaging Mission: preliminary science drivers and technical
  requirements}.
p. 99040L

\bibitem[\protect\citeauthoryear{{Nomoto}, {Kobayashi} \& {Tominaga}}{{Nomoto}
  et~al.}{2013}]{nomoto13}
{Nomoto} K.,  {Kobayashi} C.,    {Tominaga} N.,  2013, \araa, 51, 457

\bibitem[\protect\citeauthoryear{{Nomoto}, {Tominaga}, {Umeda}, {Kobayashi} \&
  {Maeda}}{{Nomoto} et~al.}{2006}]{nomoto06}
{Nomoto} K.,  {Tominaga} N.,  {Umeda} H.,  {Kobayashi} C.,    {Maeda} K.,
  2006, Nuclear Physics A, 777, 424

\bibitem[\protect\citeauthoryear{{Ohkubo}, {Nomoto}, {Umeda}, {Yoshida} \&
  {Tsuruta}}{{Ohkubo} et~al.}{2009}]{ohkubo09}
{Ohkubo} T.,  {Nomoto} K.,  {Umeda} H.,  {Yoshida} N.,    {Tsuruta} S.,  2009,
  \apj, 706, 1184

\bibitem[\protect\citeauthoryear{{Omukai}, {Tsuribe}, {Schneider} \&
  {Ferrara}}{{Omukai} et~al.}{2005}]{omukai05}
{Omukai} K.,  {Tsuribe} T.,  {Schneider} R.,    {Ferrara} A.,  2005, \apj, 626,
  627

\bibitem[\protect\citeauthoryear{{Placco} et~al.,}{{Placco}
  et~al.}{2018}]{placco18}
{Placco} V.~M.,  et~al., 2018, \aj, 155, 256

\bibitem[\protect\citeauthoryear{{Placco}, {Frebel}, {Beers}, {Yoon}, {Chiti},
  {Heger}, {Chan}, {Casey} \& {Christlieb}}{{Placco} et~al.}{2016}]{placco16}
{Placco} V.~M.,  {Frebel} A.,  {Beers} T.~C.,  {Yoon} J.,  {Chiti} A.,  {Heger}
  A.,  {Chan} C.,  {Casey} A.~R.,    {Christlieb} N.,  2016, \apj, 833, 21

\bibitem[\protect\citeauthoryear{{Placco}, {Frebel}, {Lee}, {Jacobson},
  {Beers}, {Pena}, {Chan} \& {Heger}}{{Placco} et~al.}{2015}]{placco15}
{Placco} V.~M.,  {Frebel} A.,  {Lee} Y.~S.,  {Jacobson} H.~R.,  {Beers} T.~C.,
  {Pena} J.~M.,  {Chan} C.,    {Heger} A.,  2015, \apj, 809, 136

\bibitem[\protect\citeauthoryear{{Roederer} \& {Barklem}}{{Roederer} \&
  {Barklem}}{2018}]{roederer18}
{Roederer} I.~U.,  {Barklem} P.~S.,  2018, \apj, 857, 2

\bibitem[\protect\citeauthoryear{{Roederer}, {Placco} \& {Beers}}{{Roederer}
  et~al.}{2016}]{roederer16}
{Roederer} I.~U.,  {Placco} V.~M.,    {Beers} T.~C.,  2016, \apjl, 824, L19

\bibitem[\protect\citeauthoryear{{Shen}, {Kulkarni}, {Madau} \& {Mayer}}{{Shen}
  et~al.}{2017}]{shen16}
{Shen} S.,  {Kulkarni} G.,  {Madau} P.,    {Mayer} L.,  2017, \mnras, 469, 4012

\bibitem[\protect\citeauthoryear{{Sneden}, {Cowan}, {Kobayashi}, {Pignatari},
  {Lawler}, {Den Hartog} \& {Wood}}{{Sneden} et~al.}{2016}]{sneden16}
{Sneden} C.,  {Cowan} J.~J.,  {Kobayashi} C.,  {Pignatari} M.,  {Lawler} J.~E.,
   {Den Hartog} E.~A.,    {Wood} M.~P.,  2016, \apj, 817, 53

\bibitem[\protect\citeauthoryear{{Stacy}, {Greif} \& {Bromm}}{{Stacy}
  et~al.}{2010}]{stacy10}
{Stacy} A.,  {Greif} T.~H.,    {Bromm} V.,  2010, \mnras, 403, 45

\bibitem[\protect\citeauthoryear{{Suda} et~al.,}{{Suda}  et~al.}{2008}]{saga}
{Suda} T.,  et~al., 2008, \pasj, 60, 1159

\bibitem[\protect\citeauthoryear{{Tanaka}, {Chiaki}, {Tominaga} \&
  {Susa}}{{Tanaka} et~al.}{2017}]{tanaka17}
{Tanaka} S.~J.,  {Chiaki} G.,  {Tominaga} N.,    {Susa} H.,  2017, \apj, 844,
  137

\bibitem[\protect\citeauthoryear{{Tanikawa}, {Suzuki} \& {Doi}}{{Tanikawa}
  et~al.}{2018}]{tanikawa18}
{Tanikawa} A.,  {Suzuki} T.~K.,    {Doi} Y.,  2018, \pasj, 70, 80

\bibitem[\protect\citeauthoryear{{Tominaga}}{{Tominaga}}{2009}]{tominaga09}
{Tominaga} N.,  2009, \apj, 690, 526

\bibitem[\protect\citeauthoryear{{Tominaga}, {Iwamoto} \& {Nomoto}}{{Tominaga}
  et~al.}{2014}]{tominaga14}
{Tominaga} N.,  {Iwamoto} N.,    {Nomoto} K.,  2014, \apj, 785, 98

\bibitem[\protect\citeauthoryear{{Tominaga}, {Umeda} \& {Nomoto}}{{Tominaga}
  et~al.}{2007}]{tominaga07}
{Tominaga} N.,  {Umeda} H.,    {Nomoto} K.,  2007, \apj, 660, 516

\bibitem[\protect\citeauthoryear{{Turk}, {Abel} \& {O'Shea}}{{Turk}
  et~al.}{2009}]{turk09}
{Turk} M.~J.,  {Abel} T.,    {O'Shea} B.,  2009, Science, 325, 601

\bibitem[\protect\citeauthoryear{{Umeda} \& {Nomoto}}{{Umeda} \&
  {Nomoto}}{2002}]{umeda02}
{Umeda} H.,  {Nomoto} K.,  2002, \apj, 565, 385

\bibitem[\protect\citeauthoryear{{Yoshii}}{{Yoshii}}{1981}]{yoshii81}
{Yoshii} Y.,  1981, \aap, 97, 280

\end{thebibliography}

\label{lastpage}

\end{document}